# Classifying superconductivity in an infinite-layer nickelate $Nd_{0.8}Sr_{0.2}NiO_2$

E. F. Talantsev[1,2*]

[1]M.N. Mikheev Institute of Metal Physics, Ural Branch, Russian Academy of Sciences, 18, S. Kovalevskoy St., Ekaterinburg, 620108, Russia

[2]NANOTECH Centre, Ural Federal University, 19 Mira St., Ekaterinburg, 620002, Russia

[*]E-mail: evgeny.talantsev@imp.uran.ru

*Abstract*

Recently Li *et al* (2019 Nature **572** 624) discovered a new type of oxide superconductor $Nd_{0.8}Sr_{0.2}NiO_2$ with $T_c = 14$ K. To classify superconductivity in this infinite-layer nickelate experimental upper critical field, $B_{c2}(T)$, and the self-field critical current densities, $J_c(sf,T)$, reported by Li *et al* (2019 Nature **572** 624), are analysed in assumption of *s*-, *d*-, and *p*-wave pairing symmetries and single- and multiple-band superconductivity. Based on deduced the ground-state superconducting energy gap, $\Delta(0)$, the London penetration depth, $\lambda(0)$, the relative jump in electronic specific heat at $T_c$, $\Delta C/C$, and the ratio of $2\Delta(0)/k_B T_c$, we conclude that $Nd_{0.8}Sr_{0.2}NiO_2$ is type-II high-$\kappa$ weak-coupled single-band *s*-wave superconductor.

# Classifying superconductivity in an infinite-layer nickelate $Nd_{0.8}Sr_{0.2}NiO_2$

## I. Introduction

For several decades the term of infinite-layer superconductor was referred to a copper-oxide superconducting compounds, $Sr_{1-x}M_xCuO_2$ (M=La, Nd, Ca, Sr…) [1,2], until recently, Li *et al* [3] have extended this class of unconventional superconductors by the discovery of superconductivity at $T_c = 14$ K in $Nd_{0.8}Sr_{0.2}NiO_2$ nickelate. Thus, bulk superconducting oxides family, i.e. tungsten bronzes [4], titanates [5], bismuthates [6], cuprates [7], and ruthenates [8] extends by a new nickelate member. Several research groups proposed different models for superconducting state in this compound [9-12], and the exhibiting of the superconducting state in this compound is in a debate [13].

In this paper, to classify superconductivity in this new class of oxide superconductors the temperature-dependent upper critical field, $B_{c2}(T)$, and the self-field critical current density, $J_c(sf,T)$, are analysed within *s*-, *d*-, and *p*-pairing symmetries. In result, it is shown that infinite-layer $Nd_{0.8}Sr_{0.2}NiO_2$ nickelate is weak-coupled single band *s*-wave superconductor.

## II. Models description

The Ginzburg-Landau theory [14] has two fundamental lengths, one is the coherence length, $\xi(T)$, and the second is London penetration depth, $\lambda(T)$. The ground state coherence length, $\xi(0)$, is given by [14,15]:

$$B_{c2}(0) = \frac{\phi_0}{2\cdot\pi\cdot\xi^2(0)}, \quad (1)$$

where $\phi_0 = 2.068 \cdot 10^{-15}$ Wb is magnetic flux quantum, and $B_{c2}(0)$ is the ground state upper critical field. For temperature dependent coherence length, $\xi(T)$, several models were proposed [14,15-21]. In this paper, to deduce the ground state coherence length, $\xi(0)$, in infinite-layer $Nd_{0.8}Sr_{0.2}NiO_2$ nickelate superconductor, three models are used. The first model was proposed by Gor'kov [16,17] (Gor'kov model):

$$B_{c2}(T) = \frac{\phi_0}{2\cdot\pi\cdot\xi^2(0)} \cdot \left(\frac{1.77-0.43\cdot\left(\frac{T}{T_c}\right)^2+0.07\cdot\left(\frac{T}{T_c}\right)^4}{1.77}\right) \cdot \left[1-\left(\frac{T}{T_c}\right)^2\right]. \qquad (2)$$

The second model was proposed by Baumgartner *et al* [20] (B-WHH):

$$B_{c2}(T) = \frac{\phi_0}{2\cdot\pi\cdot\xi^2(0)} \cdot \left(\frac{\left(1-\frac{T}{T_c}\right)-0.153\cdot\left(1-\frac{T}{T_c}\right)^2-0.152\cdot\left(1-\frac{T}{T_c}\right)^4}{0.693}\right) \qquad (3)$$

And the third model was proposed recently in our recent report [21]:

$$B_{c2}(T) = \frac{\phi_0}{2\cdot\pi\cdot\xi^2(0)} \cdot \left[\left(\frac{1.77-0.43\cdot\left(\frac{T}{T_c}\right)^2+0.07\cdot\left(\frac{T}{T_c}\right)^4}{1.77}\right)^2 \cdot \frac{1}{1-\frac{1}{2\cdot k_B\cdot T}\int_0^\infty \frac{d\varepsilon}{\cosh^2\left(\frac{\sqrt{\varepsilon^2+\Delta^2(T)}}{2\cdot k_B\cdot T}\right)}}\right] \qquad (4)$$

where $k_B$ is Boltzmann constant, and $\Delta(T)$ is the temperature-dependent superconducting gap, for which analytical expression was given by Gross *et al* [22]:

$$\Delta(T) = \Delta(0) \cdot \tanh\left[\frac{\pi\cdot k_B\cdot T_c}{\Delta(0)} \cdot \sqrt{\eta\cdot\frac{\Delta C}{C}\cdot\left(\frac{T_c}{T}-1\right)}\right] \qquad (5)$$

where $\Delta(0)$ is the ground state energy gap amplitude, $\Delta C/C$ is the relative jump in electronic specific heat at $T_c$, $\eta = 2/3$ for *s*-wave superconductors [22].

Thus, $\xi(0)$ and $T_c$ can be obtained by fitting experimental $B_{c2}(T)$ data to Eqs. 2-4. In addition, $\Delta C/C$, $\Delta(0)$ and, thus, the ratio of $\frac{2\Delta(0)}{k_B T_c}$, can be deduced as free-fitting parameters by fitting experimental $B_{c2}(T)$ data to Eq. 4. More details about the procedures can be found elsewhere [23].

There is an alternative way to deduce $\Delta(0)$, $\Delta C/C$, $T_c$ and $\frac{2\Delta(0)}{k_B T_c}$ by the fit of experimental self-field critical current density, $J_c(\text{sf},T)$, to universal equation, which is for thin-film superconductors reduced to simple form [23,24]:

$$J_c(\text{sf},T) = \frac{\phi_0}{4\pi\mu_0} \cdot \frac{\ln\left(1+\sqrt{2}\cdot\frac{\lambda(0)}{\xi(0)}\right)}{\lambda^3(T)} \qquad (6)$$

where $\phi_0 = 2.067 \times 10^{-15}$ Wb is the magnetic flux quantum, $\mu_0 = 4\pi \times 10^{-7}$ H/m is the magnetic permeability of free space, and the London penetration depth, $\lambda(T)$, is given by:

1. $$\lambda(T) = \frac{\lambda(0)}{\sqrt{1 - \frac{1}{2 \cdot k_B \cdot T} \cdot \int_0^\infty \frac{d\varepsilon}{\cosh^2\left(\frac{\sqrt{\varepsilon^2 + \Delta^2(T)}}{2 \cdot k_B \cdot T}\right)}}},\qquad(7)$$

for *s*-wave superconductors, where Δ(*T*) is given by Eq. 5 [22,25].

2. $$\lambda(T) = \frac{\lambda(0)}{\sqrt{1 - \frac{1}{2 \cdot k_B \cdot T} \cdot \int_0^{2\pi} \cos^2(\theta) \cdot \left(\int_0^\infty \frac{d\varepsilon}{\cosh^2\left(\frac{\sqrt{\varepsilon^2 + \Delta^2(T,\theta)}}{2 \cdot k_B \cdot T}\right)}\right) \cdot d\theta}} \qquad(8)$$

for *d*-wave superconductors, where the superconducting energy gap, Δ(*T*,θ), is given by [22,25]:

$$\Delta(T, \theta) = \Delta_m(T) \cdot \cos(2\theta) \qquad(9)$$

where Δ$_m$(*T*) is the is the maximum amplitude of the *k*-dependent *d*-wave gap given by Eq. 5, θ is the angle around the Fermi surface subtended at (π, π) in the Brillouin zone (details can be found elsewhere [22,25,26]). In Eq. 9 the value of η = 7/5 [22,25,26].

3. And *p*-wave symmetry [22,25], which only recently was tested to fit critical current densities in superconductors [22,25]:

$$\lambda_{(p,a)(\perp,\parallel)}(T) = \frac{\lambda_{(p,a)(\perp,\parallel)}(0)}{\sqrt{1 - \frac{3}{4 \cdot k_B \cdot T} \cdot \int_0^1 w_{\perp,\parallel}(x) \cdot \left(\int_0^\infty \frac{d\varepsilon}{\cosh^2\left(\frac{\sqrt{\varepsilon^2 + \Delta_{p,a}^2(T) \cdot f_{p,a}^2(x)}}{2 \cdot k_B \cdot T}\right)}\right) \cdot dx}} \qquad(10)$$

where subscripts *p*, *a*, ⊥, and ∥ designate polar, axial, perpendicular and parallel cases respectively. For this symmetry, the gap function is given by [22,25]:

$$\Delta(\hat{\boldsymbol{k}}, T) = \Delta(T) f(\hat{\boldsymbol{k}}, \hat{\boldsymbol{l}}) \qquad(11)$$

where, Δ(*T*) is the superconducting gap amplitude, ***k*** is the wave vector, and ***l*** is the gap axis. Thus, temperature dependence of λ(*T*) is determined by mutual orientation of the vector potential, ***A***, and the gap axis, ***l***, which is for transport current experiment just the orientation of the crystallographic axes of the film compared with the direction of the electric current.

There are two distinctive orientations, $A \perp l$ (when $A$ is perpendicular to $l$) and polar $A \| l$ (when $A$ is parallel to $l$) [22,25]. More details can be found elsewhere [22,25,26]). The function of $w_{\perp,\|}(x)$ in Eq. 10 is:

$$w_\perp(x) = (1 - x^2)/2 \tag{12}$$

and

$$w_\|(x) = x^2 \tag{13}$$

and the gap amplitude in Eq. 11 is just Eq. 5, but $\eta$ is given by [25]:

$$\eta_{p,a} = \frac{2}{3} \cdot \frac{1}{\int_0^1 f_{p,a}^2(x) \cdot dx} \tag{14}$$

where

$$f_p(x) = x \text{ ; polar configuration} \tag{15}$$

$$f_a(x) = \sqrt{1 - x^2} \text{ ; axial configuration} \tag{16}$$

More details about the $J_c(\text{sf},T)$ analysis for $p$-wave symmetry can be found elsewhere [26,27].

By substituting Eqs. 5, 7-13 in Eq. 6, one can fit experimental $J_c(\text{sf},T)$ data to $s$-, $d$-, $p$-wave gap symmetries to deduce $\lambda(0)$, $\Delta(0)$, $\Delta C/C$, $T_c$ and $\frac{2\Delta(0)}{k_B T_c}$ as free-fitting parameters. This approach is recently applied for wide range of thin film unconventional superconductors [23,24,26-32].

## III. $B_{c2}(T)$ analysis

There are several criteria to define $B_{c2}(T)$ from experimental $R(T)$ curves. In this paper ti define $B_{c2}(T)$ we use the criterion of 3% of normal state resistance, $R_{\text{norm}}(T)$, for $R(T)$ curves of $Nd_{0.8}Sr_{0.2}NiO_2$ presented in Fig. 4(a) by Li *et al* [3]. The fits of $B_{c2}(T)$ data to three models are shown in Fig. 1. It can be seen that $\xi(0)$ values deduced by three models are close to each other and following analysis of $J_c(\text{sf},T)$ will be utilized an average value of:

$$\xi(0) = 5.7 \pm 0.3 \; nm. \tag{17}$$

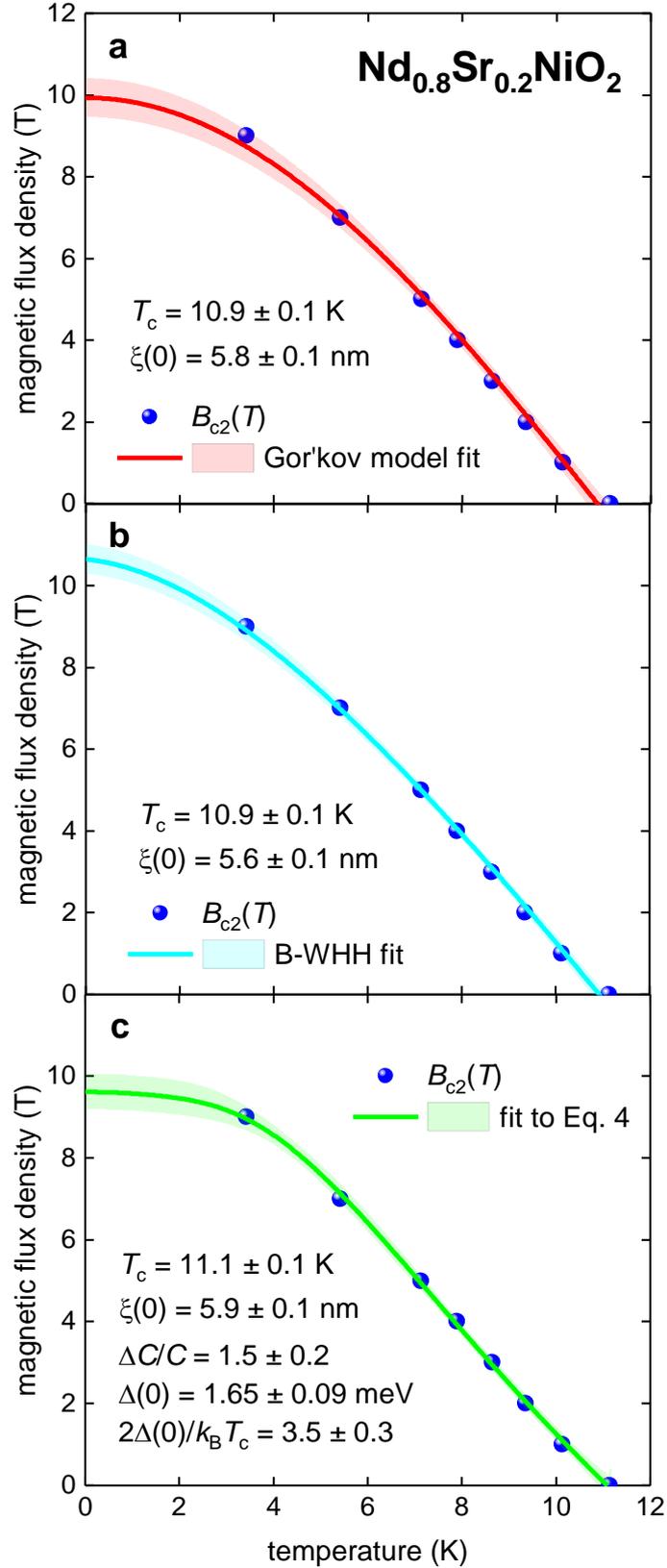

**Figure 1.** The upper critical field, $B_{c2}(T)$, of $Nd_{0.8}Sr_{0.2}NiO_2$ (reported by Li *et al* [3]) and data fits to three models (Eqs. 2-4). (a) fit to Gor'kov model, the fit quality is $R = 0.995$. (b) fit to B-WHH, $R = 0.998$. (c) fit to Eq. 4, $R = 0.9993$. 95% confidence bars are shown.

This deduced value for $\xi(0)$ is in reasonable agreement with $\xi(0) = 4.5$ nm reported by Jovanović et al. [33] for copper-oxide-based infinite layer counterpart of $La_{1-x}Sr_xCuO_2$.

Deduced values by the fit to Eq. 4:

$$\frac{2\Delta(0)}{k_B T_c} = 3.5 \pm 0.3 \tag{18}$$

$$\frac{\Delta C}{C} = 1.5 \pm 0.2 \tag{19}$$

are, within uncertainties, equal to BCS [34] weak-coupling limits of 3.53 and 1.43 respectively, and the former deduced value is equal to recently deduced value of:

$$\frac{2\Delta(0)}{k_B T_c} = 3.51 \pm 0.05 \tag{20}$$

for *s*-wave oxide superconductor of $Ba_{0.51}K_{0.49}BiO_3$ [35].

It should be noted that there is no sign in experimental $B_{c2}(T)$ data that $Nd_{0.8}Sr_{0.2}NiO_2$ exhibits two superconducting band state, which can be seen as sharp enhancement in amplitude of $B_{c2}(T)$ at critical temperature of the second superconducting band opening (see for details Ref. 36).

## IV. $J_c(sf,T)$ analysis

The critical current density, $J_c$, is defined as the lowest, detectable in experiment, value of electric power dissipation in a superconductor on electric current flow. For available $E(I)$ curves presented by Li *et al* [3] in their Fig. 3(f), the critical current density at self-field condition (when no external magnetic field is applied), $J_c(sf,T)$, can be defined at the lowest value of electric field of $E_c = 3$ V/cm. Experimental $J_c(sf,T)$ deduced by this $E_c$ criterion and the fit to single band *s*-wave model (i.e., Eqs. 6,7 for which $\xi(0) = 5.7$ nm was fixed) are shown in Fig. 2(a). It can be seen that the fit is excellent, and deduced superconducting parameters (Fig. 2(a) and Table 1) are within BCS weak-coupling limits.

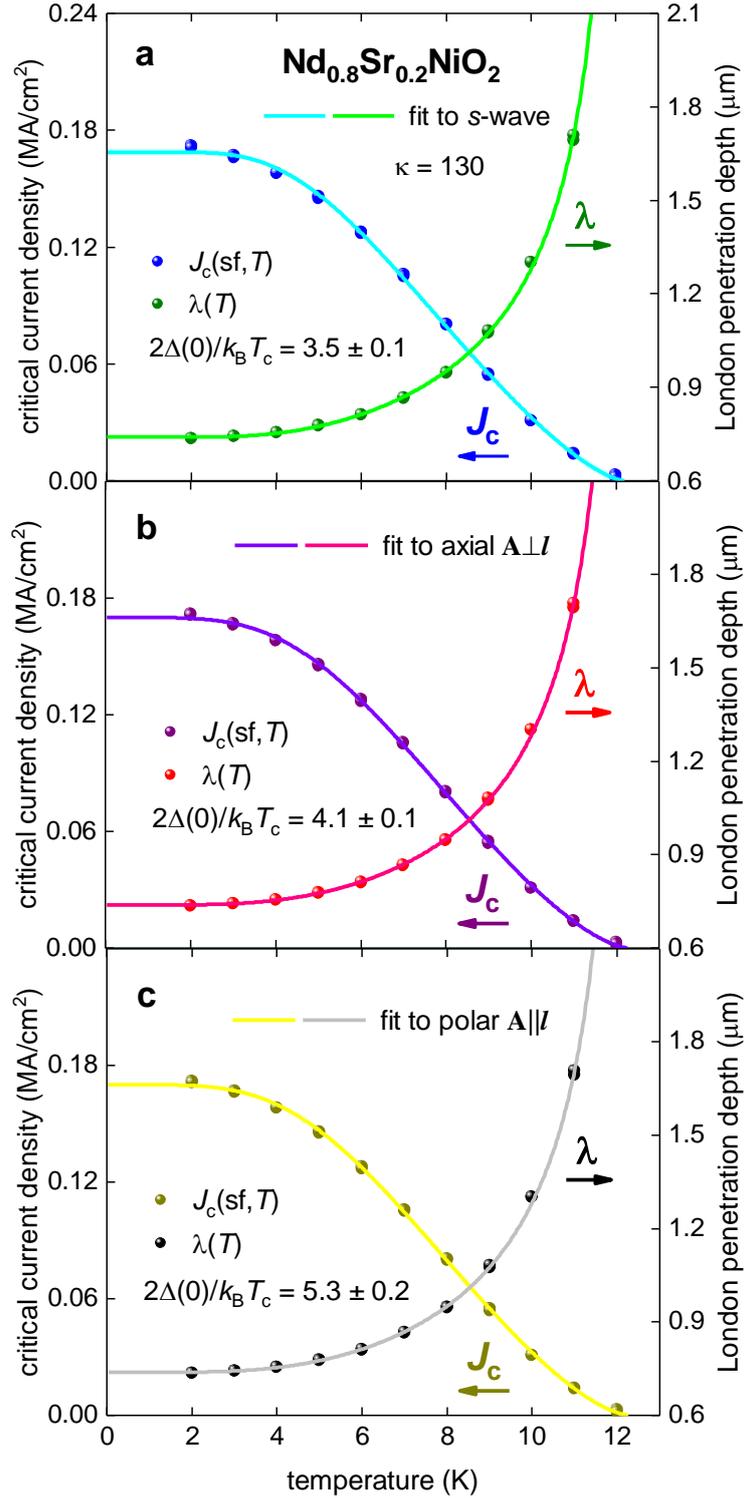

**Figure 2.** The self-field critical current density, $J_c(\text{sf},T)$, for $Nd_{0.8}Sr_{0.2}NiO_2$ thin film with raw data processed from the work of Li *et al.* [3] and a fit of the data to three single-band models. For all models $\xi(0) = 5.7$ nm was used. (a) *s*-wave fit, $\lambda(0) = 740 \pm 3$ nm, $T_c = 12.2 \pm 0.1$ K, the goodness of fit $R = 0.995$; (b) *p*-wave axial $\mathbf{A}\perp l$ fit, $\lambda(0) = 738 \pm 2$ nm, $T_c = 12.3 \pm 0.1$ K, $R = 0.997$; (c) *p*-wave polar $\mathbf{A}\|l$ fit, $\lambda(0) = 735 \pm 2$ nm, $T_c = 12.4 \pm 0.1$ K $R = 0.9990$. Other deduced parameters are listed in Table I.

Deduced $\lambda(0) = 740 \pm 3$ nm is similar to $\lambda(0) = 690\text{-}850$ nm measured for samples possessing maximal $T_c$ values for cuprate counterpart $La_{1-x}Sr_xCuO_2$ [37].

By utilizing deduced $\lambda(0)$ value the Ginzburg-Landau parameter $\kappa = \frac{\lambda(0)}{\xi(0)} = 130$ which is similar to $La_{1-x}Sr_xCuO_2$ [33,37] and this value is at the upper-level range for other cuprates and unconventional superconductors [15,23,24,26,38-43].

**Table I.** Deduced $2\Delta(0)/k_BT_c$ and $\Delta C/C$ values for $Nd_{0.8}Sr_{0.2}NiO_2$ from $J_c$(sf,$T$) fits to Eqs. 6-8 and BCS weak-coupling limits for the same parameters within for *s*-, *d*-, and *p*-wave pairing symmetries [20,23]. For *d*-wave symmetry, $\Delta_m(0)$ was used (which is the maximum amplitude of the *k*-dependent *d*-wave gap, $\Delta(\theta) = \Delta_m(0)\cos(2\theta)$).

| Pairing symmetry and experiment geometry | Deduced $\frac{2\Delta(0)}{k_BT_c}$ | BCS weak-coupling limit of $\frac{2\Delta(0)}{k_BT_c}$ | Deduced $\frac{\Delta C}{C}$ | BCS weak-coupling limit of $\frac{\Delta C}{C}$ |
|---|---|---|---|---|
| *s*-wave | $3.5 \pm 0.2$ | 3.53 | $1.5 \pm 0.2$ | 1.43 |
| *d*-wave | $> 10^2$ | 4.28 | $2.3 \pm 0.5$ | 0.995 |
| *p*-wave; axial $\mathbf{A}\perp \mathbf{l}$ | $4.1 \pm 0.1$ | 4.06 | $1.07 \pm 0.08$ | 1.19 |
| *p*-wave; axial $\mathbf{A}\|\mathbf{l}$ | $9.0 \pm 2.4$ | 4.06 | $1.55 \pm 0.05$ | 1.19 |
| *p*-wave; polar $\mathbf{A}\perp \mathbf{l}$ | $> 5\cdot 10^2$ | 4.92 | $2.5 \pm 0.4$ | 0.79 |
| *p*-wave; polar $\mathbf{A}\|\mathbf{l}$ | $5.3 \pm 0.3$ | 4.92 | $0.63 \pm 0.03$ | 0.79 |

Alternatively, *d*- and *p*-wave superconducting gap symmetries can be considered. The fits to *d*-wave symmetry, as well as to polar $\mathbf{A}\perp \mathbf{l}$ and axial $\mathbf{A}\|\mathbf{l}$ of *p*-wave, reveal very large $\frac{2\Delta(0)}{k_BT_c}$ values and these symmetries can be excluded from further consideration.

The cases of polar $\mathbf{A}\|\mathbf{l}$ and axial $\mathbf{A}\perp \mathbf{l}$ gap symmetries are still hypothetically possible (Table 1), and $J_c$(sf,$T$) fit to these models are shown in Figs. 2(b,c) respectively, however, for given experimental conditions (i.e. epitaxial *c*-axis oriented thin film) expected geometry is polar **A**⊥*l* [21].

It should be also noted that there is no sign for two-band superconductivity in $Nd_{0.8}Sr_{0.2}NiO_2$ which usually can be detected by a sharp enhancement in $J_c(sf,T)$ at critical temperature of the second band opening [24,36].

By taking in account a good agreement between $\frac{2\Delta(0)}{k_B T_c}$ and $\frac{\Delta C}{C}$ values deduced for *s*-wave symmetry from $B_{c2}(T)$ and $J_c(sf,T)$ analyses (Eqs. 17, 18 and Table 1, respectively), which are, in addition, within BCS weak-coupling limits for this symmetry, and a fact that *s*-wave pairing symmetry is the most conventional one, we can conclude that $Nd_{0.8}Sr_{0.2}NiO_2$ nickelate is weak-coupling single band high-κ *s*-wave superconductor.

## V. Conclusions

Recently discovered [3] an infinite-layer nickelate $Nd_{0.8}Sr_{0.2}NiO_2$ superconductor is a new member of bulk oxide superconductors for which experimental $B_{c2}(T)$ and $Jc(sf,T)$ data are analysed in this paper.

In result, it is found that an infinite-layer nickelate $Nd_{0.8}Sr_{0.2}NiO_2$ is weak-coupling single band high-κ *s*-wave superconductor.

**Acknowledgement**

Author thanks Dr. W. P. Crump (Aalto University) for invaluable help, and Prof. O. P. Sushkov (University of New South Wales), Prof. P. Bourges (Universite Paris-Sacray), and Prof. G. Seibold (Brandenburgische Technische Universität Cottbus–Senftenberg) for fruitful discussions.

Author also thanks financial support provided by the state assignment of Minobrnauki of Russia (theme "Pressure" No. AAAA-A18-118020190104-3) and by Act 211 Government of the Russian Federation, contract No. 02.A03.21.0006.